\documentclass[a4paper,12pt]{article}
\usepackage[dvipsnames]{xcolor}
\usepackage[utf8]{inputenc} 
\usepackage[hidelinks]{hyperref}
\usepackage{float}
\usepackage{amssymb,amsmath,amsthm}
\usepackage{amsfonts}

\usepackage[draft,colorinlistoftodos, textwidth=3cm, textsize=scriptsize]{todonotes}

\pdfoutput=1
\usepackage{graphicx}
\usepackage{graphicx, float, tabularx, booktabs}
\usepackage{color}
\usepackage{enumerate}
\usepackage[american]{babel}
\usepackage{stackrel}

\usepackage[numbers, compress]{natbib}
\usepackage{natbib}
\usepackage{slashed}
\usepackage{amsfonts}
\usepackage{epsfig}
\usepackage{graphicx, tabularx}
\newcommand{\MP}{M_\mathrm{P}}
\newcommand{\be}{\begin{equation}}
\newcommand{\ee}{\end{equation}}
\newcommand{\ba}{\begin{eqnarray}}
\newcommand{\ea}{\end{eqnarray}}

\vspace{-0.1cm}
\renewcommand{\thefootnote}{\fnsymbol{footnote}}

\begin{document}
%\listoftodos
%\newpage

\begin{center}
{\Large \textbf{Finite electrodynamics from  T-duality}}
\end{center}

\vspace{-0.1cm}

\begin{center}
Patricio Gaete$^{a}$\footnote{%
E-mail: \texttt{patricio.gaete@usm.cl} },   and Piero Nicolini$^{b,c,d}$\footnote{%
E-mail: \texttt{nicolini@fias.uni-frankfurt.de} }

\vspace{.6truecm}

\emph{\small  $^a$Departamento de F\'{i}sica and\\[-0.5ex] Centro Cient\'{i}fico-Tecnol\'ogico de Valpara\'{i}so-CCTVal,\\[-0.5ex]
\small Universidad T\'{e}cnica Federico Santa Mar\'{i}a, Valpara\'{i}so, Chile}\\[1ex]

\emph{\small  $^b$Center for Astro, Particle and Planetary Physics, \\[-0.5ex]
\small New York University Abu Dhabi, % \\[-0.5ex]
\small  Abu Dhabi, UAE}\\[1ex] 

\emph{\small  $^c$Frankfurt Institute for Advanced Studies (FIAS),\\[-0.5ex]   Frankfurt am Main, Germany}\\[1ex]

\emph{\small  $^d$Institut f\"ur Theoretische Physik,\\[-0.5ex]  
Goethe-Universit\"at, Frankfurt am Main, Germany}\\[1ex]

%\emph{\small  $^e$Dipartimento di Ingegneria e Scienze Applicate,\\[-0.5ex]
%\small Università degli Studi di Bergamo, Dalmine (Bergamo), Italy }\\[1ex]

\end{center}
\begin{abstract}
\noindent{\small  % \noindent
In this paper, we present the repercussions of Padmanabhan's propagator in electrodynamics.
This corresponds to implement T-duality effects in a U(1) gauge theory. By formulating a
nonlocal action consistent with the above hypothesis, we derive the profile of static
potentials between electric charges via a path integral approach. Interestingly, the
Coulomb potential results regularized by a length scale proportional to the parameter 
$(\alpha^\prime)^{1/2}$. Accordingly, fields are vanishing at the origin. We also discuss an array of
experimental testbeds to expose  the above results. It is interesting to observe that T-duality generates
an effect of dimensional fractalization, that resembles similar phenomena in fractional electromagnetism. Finally, our results have also been derived  with a gauge-invariant method, as a necessary check of consistency for any non-Maxwellian theory. }
\end{abstract}

\renewcommand{\thefootnote}{\arabic{footnote}} \setcounter{footnote}{0}
%\thispagestyle{empty}
%\clearpage
%\tableofcontents

\section{Introduction}

The formulation of a unified description of  fundamental interactions governing the Universe from microscopic to macroscopic
scales represents a path fraught with many difficulties. For instance, the combination of gravity and quantum mechanics is
technically and conceptually challenging,  the origin of dark sectors and, more in general, the phenomenology beyond the 
Standard Model and Einstein gravity are not yet clear. Even more problematic is the inclusion of such a  phenomenology within a 
unique formulation, that works from the Planck  scale, $\MP\sim 10^{19}$ GeV, down to the Terascale. 

Superstring theory is probably the major contender for a self consistent description of all interactions at quantum level 
\cite{Nicolai13}. It enjoys ultraviolet finiteness \cite{ACV87,ACV88} and absence of anomalies \cite{GrS84}. It is, however, not 
free from limitations.  For instance, it is a hard tasks to identify the string landscape, namely  effective field theories that 
are equipped with genuine stringy effects \cite{Vaf05}. In addition,   superstring theory is currently in conflict with 
experimental evidence due to the non observation of supersymmetry signals\footnote{For a balanced report of the current status 
of quantum gravity proposals (including string theory) we recommend section 7 of Nicolai's paper \cite{Nicolai13}.}. It is 
probably wiser to postpone the analysis of all these problems and try to better understand the repercussions of some model 
independent features, i.e., universal characteristics common to  any formulation of quantum gravity. For instance, a quantum 
spacetime is expected to display an  intrinsic non-local character and be singularity free. As a result, a lot of efforts has 
been devoted to improve classical spacetime geometries, by following  a variety of quantum gravity paradigms, e.g.,  
noncommutative geometry \cite{NSS06,Nic09}, the generalized uncertainty principle \cite{IMN13,CMN15,KKM+19,CMMN20}, 
asymptotically safe gravity \cite{BoR00}, loop quantum gravity \cite{Mod04}, gravity self-completeness \cite{DvG13,NiS14,Nic18} 
and non-local gravity \cite{MMN11,CMN14}.

Recently, a regular black hole solution has been derived by implementing T-duality effects \cite{NSW19}. 
%The latter naturally 
%show up, when the graviton propagator has an exponential suppression at large momenta with a profile $\sim e^{-l_0k}$, that has been introduced by Padmanabhan on model independent arguments in \cite{Pad97}. 
The essence of T-duality is the invariance of string theory under inversion of the compactification 
radius, $R\longleftrightarrow \alpha^\prime/R$ \cite{GPR94}. This necessarily implies the existence of a minimal 
compactification radius $l_0\sim \sqrt{\alpha^\prime}$, namely a scale at which the classical notion of spacetime ceases to exist. 
Such a feature of spacetime is reflected also in another byproduct of string theory, namely a modification of uncertainty 
relations
\ba
\Delta x\simeq \frac{1}{\Delta p}+\alpha^\prime \Delta p
\label{eq:gup}
\ea
known as generalized uncertainty principle \cite{Mag93,Gar95,Kem95}. Eq. \eqref{eq:gup} has an additional term that prevents 
Compton wavelengths to be smaller than $\sqrt{\alpha^\prime}$. The same conclusion descends from noncommutative geometry, a 
feature of a quantum spacetime to which T-duality is connected both in the case of open and closed strings \cite{SeW99,Lus10}.

Similarly Padmanabhan's formulation of T-duality is based on a modification of the path integral representation of a field propagator \cite{Pad97}.
 To  accommodate a spacetime ``zero-point length'', paths smaller or larger than such a length should contribute in the same way 
 to the path integral. As a result, the latter has to be invariant with respect to the duality transformation 
 $s\longleftrightarrow l_0^2/s$, where $s$ is the path proper time\footnote{We incorporate factors of the ``zero-point length'' 
 by a suitable 
redefinition of $l_0$. For more details see \cite{Pad98}.}. A propagator of this kind actually reads \cite{Pad97}
\ba
G(k)=- \frac{l_0}{\sqrt{k^2}}K_1\left(l_0\sqrt{k^2}\right),
\label{eq:dualpropagator}
\ea
where $K_1(x)$ is a modified Bessel function of the second kind. 
Such a 
function introduces an exponential decaying term for large arguments, that guarantees the  ultraviolet finiteness. 
Interestingly, the exact form of Padmanabhan's propagator \eqref{eq:dualpropagator} can analytically be derived from string theory, as it has been 
shown in a series of papers \cite{SSP03,SpF05,FSP06}.

It is important to stress that the regularity of the black hole solution  \cite{NSW19} discloses another physical meaning of 
$l_0$. From the metric coefficient
\ba
g_{00}=1-\frac{2Mr^2}{\left(r^2+l_0^2\right)^{3/2}}
\label{eq:tmetric}
\ea
one realizes that the spacetime coincides with Bardeen regular black hole, after switching, the magnetic charge with the 
zero-point length, $g\to l_0$. This means that the ultraviolet regulator can be seen as the focal point of another duality, namely 
that between gravity and gauge theories\footnote{For an expanded interpretation of the dualities emerging from the metric 
\eqref{eq:tmetric},  see the analysis in \cite{EKM20} based on the double copy mechanism.}. 

On these premises, it is now imperative to explore the role of the propagator \eqref{eq:dualpropagator} for the physics of other interactions.  
Among the motivations for such an investigation, there is the fact that, in general,  not only gravity but also customary 
field theories are subject to non-local corrections at a certain scale, where point-like objects become ill defined. Within such 
a paradigm, we start in the present paper from the simplest case of a U(1) theory, namely electrodynamics.  
To this purpose, it is important to clarify the regime we aim to explore. High energy electrodynamics is expected to depart from 
 Maxwell's formulation. Strong fields show  non-linearity, a feature that offers a rich phenomenology, e.g. 
birefringence and photon-photon scattering effects \cite{Adl71},  currently under experimental scrutiny \cite{Akh02,PVLAS06}. On 
the other hand,  we will not consider such non-linear effects but rather we will stay below the so called  Schwinger limit, 
namely
\begin{eqnarray}
E&\ll & E_{\text{c}}={\frac {m_{\text{e}}^{2}c^{3}}{q_{\text{e}}\hbar }}\simeq 1.32\times 10^{18}\,\mathrm {V} /\mathrm {m} 
\label{eq:SchwingerE} \\
B& \ll & B_{\text{c}}={\frac {m_{\text{e}}^{2}c^{2}}{q_{\text{e}}\hbar }}\simeq 4.41\times 10^{9}\,\mathrm {T}.
\end{eqnarray}
Our approach aims to analyze the short scale behavior of low intensity fields. More importantly, it offers the advantage 
of understanding the net effect of T-duality without interference from other phenomena of the strong field regime. Finally, 
we will consider only static conditions throughout the paper. As one can see, this assumption will be enough to capture the 
essential features of the modified electrodynamics. 

The papers is organized as follows: in Sec. \ref{sec:3plus1} we present the derivation of the static forces between two sources 
by employing a path integral approach; in Sec. \ref{sec:disc} we discuss the significance of our findings and their phenomenological repercussions; in Sec. \ref{sec:gaugeinv} we propose an alternative derivation of static forces based on a path dependent, gauge invariant formalism; finally in Sec. \ref{sec:concl}, we summarize our results and we draw the conclusions.

\section{Finite Maxwell-like electrodynamics}
\label{sec:3plus1}

\iffalse
We turn now to the problem of obtaining the interaction energy
between static point-like sources for the model under consideration.
To do this, we shall compute the expectation value of the energy operator $H$
in the physical state $|\Phi\rangle$, which we will denote by ${\langle
H\rangle}_\Phi$. However, before going into the axionic electrodynamics, 
we would like to first consider the Maxwell-like case. This would not only provide the
theoretical setup for our subsequent work, but also fix the notation. 

\subsection{($3+1$)-D Maxwell-like case}
\fi

In this section, we aim to derive the interaction energy between static-point like sources, by assuming that photon exchange is governed by the T-dual propagator \eqref{eq:dualpropagator}. The result will turn to be equivalent to that of standard (local) electrodynamics between T-duality modified (i.e. non-local) static sources.

The starting point of our discussion is provided by the following four dimensional spacetime Lagrangian density\footnote{We 
consider natural units from this point on, namely $\hbar=c=1$.} 
\begin{equation}
\mathcal{L} =
- \frac{1}{4}F_{\mu \nu }\, {\cal O} F^{\mu \nu }, \label{NLMaxwell05}
\end{equation}
where $\cal{O}$ is the non-local operator
\begin{equation}
\mathcal{O} = {\left[ {{l_0}\sqrt \Delta\,\,  {K_1}\left( {{l_0}\sqrt \Delta  } \right)} \right]^{ - 1}}, \label{NLMaxwell10}
\end{equation}
with $\Delta  \equiv {\partial _\mu }{\partial ^\mu }$. Here $K_{1}$ is the modified Bessel function of the second kind and 
$l_{0}$ denotes a minimal length, or zero-point length of spacetime. We stress that the Lagrangian 
(\ref{NLMaxwell05}) is manifestly gauge invariant, despite the presence of a mass term $\sim 1/\l_0$. For further clarifications about this point, see Sec. \ref{sec:gaugeinv}. 

The novelty of \eqref{NLMaxwell05} is connected to the properties of the non-local operator $\mathcal{O} $. For  small momenta, the above action becomes the usual Maxwell action, namely
\ba
\mathcal{O}\approx 1 \quad \mathrm{for} \quad  l_0\sqrt \Delta \ll 1.
\ea
On the other hand, for large momenta the operator reads: 
\ba
\mathcal{O}\approx  \ \sqrt{\frac{2}{\pi}}\ e^{l_0\sqrt{\Delta}} \quad \mathrm{for} \quad l_0\sqrt \Delta \gg 1.
\ea
The above relation is the key element for having an improvement of the short distance behavior of the field strength $F_{\mu \nu }$.

\iffalse
Having characterized the new effective Lagrangian, we will work out the interaction energy through two different methods. The 
first approach is based on the usual path-integral formalism,
whereas the second one is carried out using the gauge-invariant, but path-dependent, variables formalism. 
\fi

According to the customary path integral formulation, one can write the functional generator of the Green's functions as
\begin{equation}
Z\left[ J \right] = \exp \left( { - \frac{i}{2}\int {d^4 xd^4 y}
J^\mu \left( x \right)D_{\mu \nu } \left( {x,y} \right)J^\nu  \left(
y \right)} \right), \label{NLMaxwell15}
\end{equation}
where $D_{\mu \nu } \left( {x} \right) = \int {\frac{{d^4
k}}{{\left( {2\pi } \right)^4 }}} D_{\mu \nu } \left( k \right)e^{ -
ikx}$ is the propagator in the Feynman gauge, which reads 
\begin{equation}
{D_{\mu \nu }}\left( k \right) =  - \frac{1}{{{k^2}}}\left\{ {{{\cal O}^{ - 1}}\left( {{k^2}} \right){\eta _{\mu \nu }} + \left( 
{1 - {{\cal O}^{ - 1}}\left( {{k^2}} \right)} \right)\frac{{{k_\mu }{k_\nu }}}{{{k^2}}}} \right\}. \label{NLMaxwell20}
\end{equation}
With the aid of the expression, $Z = e^{iW\left[ J \right]}$, and  (\ref {NLMaxwell15}), $W\left[ J \right]$ becomes
\begin{eqnarray}
W\left[ J \right] =  - \frac{1}{2}\int {\frac{{{d^4}k}}{{{{\left( {2\pi } \right)}^4}}}J_\mu ^ * \left( k \right)} \left[ { - 
\frac{1}{{{k^2}}}{{\cal O}^{ - 1}}\left( {{k^2}} \right){\eta _{\mu \nu }}} \right]{J_\nu }\left( k \right) \nonumber \\
 - \frac{1}{2}\int {\frac{{{d^4}k}}{{{{\left( {2\pi } \right)}^4}}}J_\mu ^ * \left( k \right)} \left[ { - \frac{1}{{{k^2}}}
 \left( {1 - {{\cal O}^{ - 1}}\left( {{k^2}} \right)} \right)\frac{{{k_\mu }{k_\nu }}}{{{k^2}}}} \right]{J_\nu }\left( k 
 \right). \nonumber\\
\label{NLMaxwell25}
\end{eqnarray}
Now, bearing in mind that the external current $J^\mu (k)$ is conserved, we promptly obtain
\begin{equation}
W\left[ J \right] = \frac{1}{2}\int {\frac{{{d^4}k}}{{{{\left( {2\pi } \right)}^4}}}} J_\mu ^ * \left( k \right)\left[ 
{\frac{{{l_0}\sqrt { - {k^2}} {K_1}\left( {{l_0}\sqrt { - {k^2}} } \right)}}{{{k^2}}}} \right]{J^\mu }\left( k \right). 
\label{NLMaxwell30}
\end{equation}
For $J_\mu  \left( x \right) \!= \!\!\left[ {Q\delta ^{\left( 3
\right)} \left( {{\bf x} - {\bf x}^{\left( 1 \right)} } \right) + Q^
\prime  \delta ^{\left( 3 \right)} \left( {{\bf x} - {\bf x}^{\left(
2 \right)} } \right)} \right] \! \delta _\mu ^0$, we find that the interaction
energy can be brought to the form
\begin{equation}
V = Q{Q^ \prime }\int {\frac{{{d^3}k}}{{{{\left( {2\pi } \right)}^3}}}} \frac{{l_{0}\sqrt {{{\bf k}^2}} }}{{{{\bf k}^2}}}{K_1}
\left( {{l_0}\sqrt {{{\bf k}^2}} } \right){e^{i{\bf k} \cdot {\bf r}}}, \label{NLMaxwell35}
\end{equation}
where $ {\bf r} \equiv {\bf x}^{\left( 1 \right)}  - {\bf x}^{\left(2 \right)}$. 
By using the integral representation of the Bessel function (see, $3.471 (12)$ of \cite{Gradsh})
\begin{equation}
\int_0^\infty  {{x^{\nu  - 1}}\exp \left( { - x - \frac{{{\mu ^2}}}{{4x}}} \right)} dx = 2{\left( {\frac{\mu }{2}} \right)^\nu }
{K_\nu }\left( \mu  \right),  \label{NLMaxwell35a}
\end{equation}
we express the integral over ${\bf k}$ in the form
\begin{eqnarray}
&&\int {\frac{{{d^3}k}}{{{{\left( {2\pi } \right)}^3}}}} \frac{{{l_0}\sqrt {{{\bf k}^2}} {K_1}\left( {{l_0}\sqrt {{{\bf k}^2}} } 
\right)}}{{{{\bf k}^2}}}{e^{i{\bf k}\cdot {\bf r}}} = 
%\nonumber\\ &&=
\int_0^\infty  {dx}\,\, {e^{ - x}}\int {\frac{{{d^3}k}}{{{{\left( {2\pi } \right)}^3}}}} \frac{{{e^{ - \frac{{l_0^2{{\bf k}^2}}}
{{4x}}}}}}{{{{\bf k}^2}}}{e^{i{\bf k} \cdot {\bf r}}} =  \nonumber\\
&&=\frac{1}{{4{{\left( \pi  \right)}^{{\textstyle{3 \over 2}}}}}}\frac{1}{r}\int_0^\infty  {{e^{ - x}}} \gamma \left( 
{{\raise0.7ex\hbox{$1$} \!\mathord{\left/
 {\vphantom {1 2}}\right.\kern-\nulldelimiterspace}
\!\lower0.7ex\hbox{$2$}},\frac{{x{r^2}}}{{l_0^2}}} \right)
 = \frac{1}{{4\pi }}\frac{1}{{\sqrt {{r^2} + l_0^2} }}. \label{NLMaxwell40}
\end{eqnarray}
with $r \!=\! |{\bf r}|$, $k \!=\! |{\bf k}|$. Here $\gamma \left( {{\raise0.7ex\hbox{$1$} \!\mathord{\left/
 {\vphantom {1 2}}\right.\kern-\nulldelimiterspace}
\!\lower0.7ex\hbox{$2$}},\frac{{x\,{r^2}}}{{l_0^2}}} \right)$ is the lower incomplete Gamma function defined
by the following integral representation
\begin{equation}
\gamma \left( {\frac{a}{b},x} \right) \equiv \int_0^x {\frac{{du}}{u}} \, {u^{{\raise0.5ex\hbox{$\scriptstyle a$}
\kern-0.1em/\kern-0.15em
\lower0.25ex\hbox{$\scriptstyle \,b$}}}}\,{e^{ - u}}. \label{NLMaxwell40-a}
\end{equation} 

Combining (\ref{NLMaxwell35}) and (\ref{NLMaxwell40}), together with $Q^\prime= -\,Q$, the interaction energy reads
\begin{equation}
V\left( r \right) = - \frac{{{Q^2}}}{{4\pi }}\frac{1}{{\sqrt {{r^2} + l_0^2} }}. \label{NLMaxwell45}
\end{equation}
In contrast to the usual Maxwell's theory, the above energy is regular at the origin. This is the direct consequence of the string T-duality encoded in the propagator  (\ref{NLMaxwell30}), in analogy to what found for the Newton's potential in \cite{NSW19}.

One can further observe that  (\ref{NLMaxwell40}) is nothing but the familiar Green's function, $\tilde G\left( {{\bf z},{{\bf 
z}^ \prime }} \right)$, for the T-duality modified electrodynamics, namely,
\begin{equation}
{\nabla ^2}\tilde G\left( {{\bf z},{{\bf z}^ \prime }} \right) =  - {l_0} \sqrt { - {\nabla ^2}}\, {K_1}\left( {{l_0}\sqrt { - 
{\nabla ^2}} } \right){\delta ^{\left( 3 \right)}}\left( {{\bf z} - {{\bf z}^ \prime }} \right).   \label{NLMaxwell45a}
\end{equation}
where the standard (i.e. Maxwell theory) Green's function, $G_0 \left( {{\bf z},{{\bf z}^ \prime }} \right)$, reads
\begin{equation}
G_0 \left( {{\bf z},{{\bf z}^ \prime }} \right)={\left[ {{l_0}\sqrt{- {\nabla ^2}}\,\,  {K_1}\left( {{l_0}\sqrt{ - {\nabla ^2}}  
} \right)} \right]^{ - 1}}\tilde G\left( {{\bf z},{{\bf z}^ \prime }} \right). 
\end{equation}
In fact, by transforming $\tilde G\left( {{\bf z},{{\bf z}^ \prime }} \right)$  into Fourier space, we obtain 
(\ref{NLMaxwell40}).
The equation for the Green's function in \eqref{NLMaxwell45a} displays a non-local smearing of the source, that resembles what 
found in the context of noncommutative geometry \cite{SSN06,KoN10}.

Alternatively one can write \eqref{NLMaxwell45a} as
\ba
{\cal D }^2\tilde G\left( {{\bf z},{{\bf z}^ \prime }} \right) =  - {\delta ^{\left( 3 \right)}}\left( {{\bf z} - {{\bf z}^ \prime }} \right),
\label{eq:nonlocgreen}
\ea
with %\\ \quad \mathrm{with}\quad   
\ba  {\cal D }^2=\frac{{\nabla ^2}}{{{l_0}\sqrt{- {\nabla ^2}}\,\,  {K_1}\left( {{l_0}\sqrt{ - {\nabla ^2}}  
} \right)} }.
\ea
Eq. \eqref{eq:nonlocgreen} is the equation of a Green's function with a non-local operator and point-like source term. It is the equivalent of \eqref{NLMaxwell45a}, that is the equation of a Green's function with a local operator and a non-local source term.

\section{Discussion and significance}
\label{sec:disc}

We present in this section some comments about the interpretation and the phenomenological repercussions of  the our results. 

The simplest Coulomb system is the hydrogen atom, that has recently been proposed as a testbed for exotic quantum field theory 
effects \cite{WNB16}. In the case of unparticles, the accuracy of measurements of hydrogen energy levels provides  limits for the unparticle scale $\Lambda_\mathcal{U}$, that can compete with those from the muon anomaly  
\cite{CKWT07} and other experiments at the LHC \cite{ABST17}. This is particularly the case for small scaling dimension $d_
\mathcal{U}\approx 1$, (i.e., in the Maxwell limit),  since unparticle mediated electrodynamics is not analytic in this limit 
and corrections become increasingly large.  For the potential \eqref{NLMaxwell45}, the study of effects on the hydrogen atom  
has been considered in \cite{WoB19}. In this case, however, the constraints are less effective since the theory depends on one 
parameter only, $l_0$, and it is analytic for $l_0\to 0$. One can estimate this also by considering
\ba
\frac{\Delta V(a_0)}{V_0(a_0)}\approx \frac{1}{2}\frac{l_0^2}{a_0^2}<\epsilon 
\ea
where $a_0\simeq 5.29\times 10^{-11}$ m is the Bohr radius, $\Delta V(a_0)=V(a_0)-V_0(a_0)$, with $V_0(r)$ the Coulomb potential 
energy. Here $\epsilon\simeq 7.35\times 10^{-10}$ is the relative error of standard electrodynamics for the ground state energy, 
namely the uncertainty on the Rydberg \cite{CODATA21,Kra10}. Accordingly, one finds that $l_0<2.03\times 10^{-15}$m, in 
agreement with the full analysis in \cite{WoB19}.

Another possible testbed for electrodynamics is offered by the Casimir effect. In case of unparticle modifications, one can  
find strong bounds on parameters governing the theory, because of the aforementioned singularity of the Maxwell limit 
 \cite{FNP17}. 
Again, the T-duality electrodynamics is not expected to be strongly constrained in Casimir systems. The only possibility would 
be to postulate that the action \eqref{NLMaxwell05}, rather than being an extension of standard electrodynamics, describes 
another sector beyond the Standard Model, e.g., a magnetic 
monopole as suggested by the role of $l_0$ in the the metric \eqref{eq:tmetric}.
Accordingly, the Maxwell limit would turn to be singular in analogy to the unparticle case.

Interestingly, in the presence of unparticles, one expects the full Casimir plate fractalization. In other words, an un-photon 
field between the capacitor plates ``sees'' their dimension as a continuous number, rather than simply two dimensions
 for a plane. 
The same phenomenon occurs in the case of the event horizon of a black hole in  un-particle mediated gravity \cite{GHS10}. 
This can be explained by the conformal flatness of two 
dimensional spacetimes and by the fact that unparticles enjoy scale invariance \cite{NiS11}. Even if this is not the case for 
T-dual electrodynamics, the fractalizazion does show up in \eqref{NLMaxwell45} too, since the spacetime dimension becomes a 
continuous number depending on $l_0$. Indeed one can define the following fractal dimension
\ba
\mathbb{D}\equiv 3-\frac{\partial\ln V(r)}{\partial\ln r}
\ea
corresponding to the power law of a generic static potential $V\sim 1/r^{\mathbb{D}-3}$. For the Coulomb case, the above 
equation gives $\mathbb{D}=4$, signaling the absence of fractalization and the absence of non-local effects, as expected. 
Conversely for the T-duality case one finds the following scale dependent dimension:
\ba
\mathbb{D}=3+\frac{r^2}{r^2+l_0^2}.
\ea
The above equation discloses two regimes: a Coulomb regime corresponding to the standard infrared dimension $\mathbb{D}=4$ at $r
\gg l_0$, and a ultraviolet dimensionally reduced regime, $\mathbb{D}\approx 3.5$ for $r\sim l_0$. The domain of the above 
dimension is actually $3\leq \mathbb{D} < 4$. Notice that $\mathbb{D}=3$ means that the potential is constant at the origin, $V\sim 1/r^{0}$,  namely the signature of a finite electrodynamics. 

\begin{figure}[h]
\begin{center}
\includegraphics[width=0.7\textwidth]{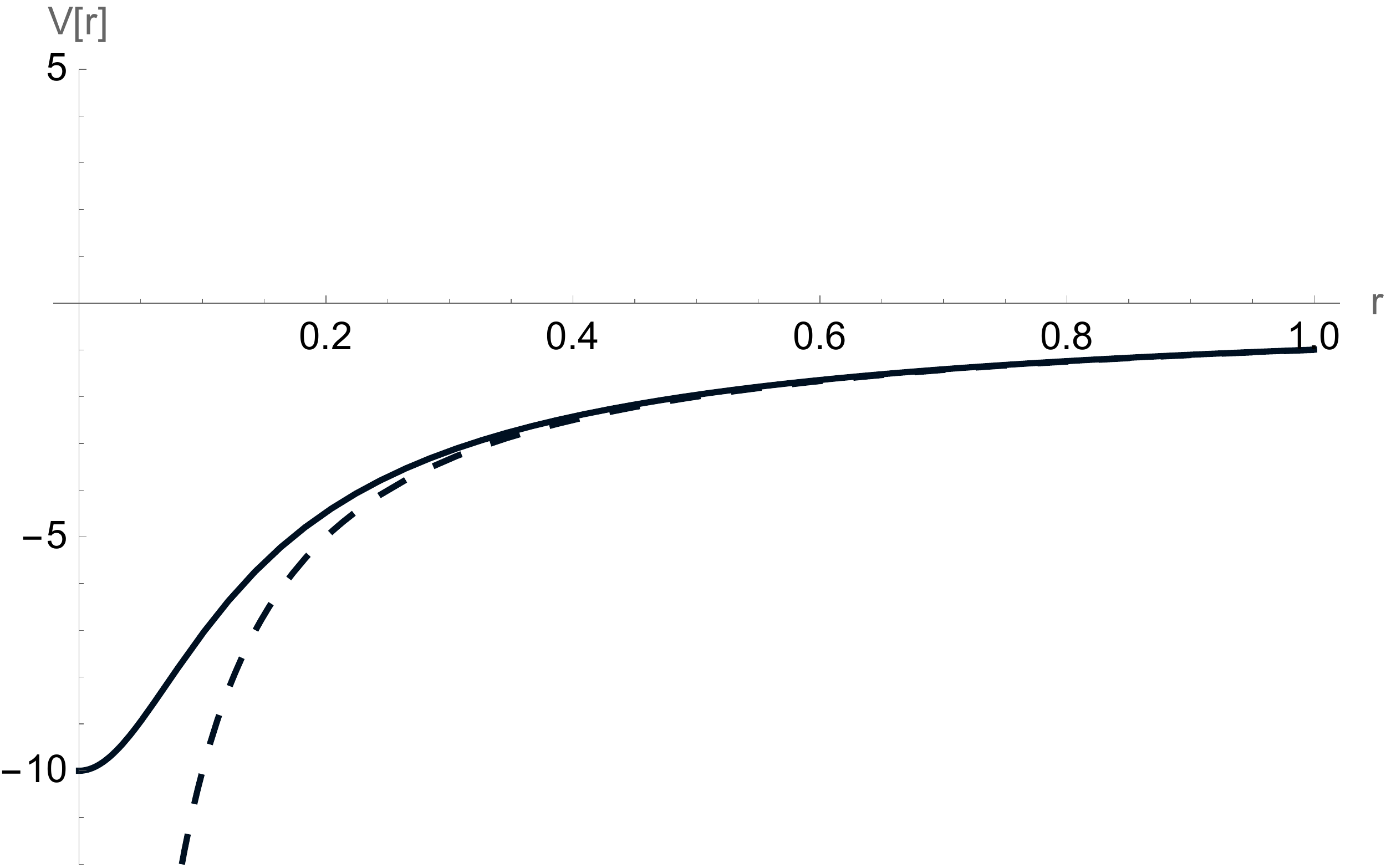}
\end{center}
\caption{\small
Shape of the potential energy (\ref{NLMaxwell45}) for  $\frac{{{Q^2}}}{4\pi}=1$ and $l_0=0.1$. The dashed line represents 
the Coulomb potential.}
\label{fig1}
\end{figure}

At this points, a remark about the short distance limits, $r\sim l_0$ and $r\ll l_0$,  is mandatory. At first sight, it could appear counter-intuitive to consider length scales below the minimal length $l_0$. Ideally, one would like to perform a physical limit, $r\to\l_0$, to obtain a finite potential $V\sim 1/l_0$. Such a physical limit would require a ``Swiss cheese'' spacetime structure. In practice, cutting off a  ``ball'' centered  around the charge out of the manifold  is mathematically hard or impossible to do. To circumvent such a difficulty,  in our formulation we traded the cut off of the manifold ball with an integrand function in \eqref{NLMaxwell40} that provides negligible contributions to the electrostatic potential for $r<l_0$ (see Fig \ref{fig1}). As a result, it is formally still possible to consider limits of the fractal dimension for $r\to 0$. Such a limit tells us that the theory is finite.  The other case, namely $r\sim l_0$, tells us that in such a regime the manifold is already subjected to wild quantum mechanically fluctuations. The spacetime is actually a fractal with non-integer dimension. 

From  \eqref{NLMaxwell45}, one can derive the electric field $\mathbf{E}=-\nabla V/Q$, where $V/Q$ is the electric  potential. 
After taking the gradient one finds:
\ba
\mathbf{E}(\mathbf{r})= -\frac{Q\ r}{4\pi \left(r^2+l_0^2\right)^{3/2}}\ \hat{\textbf{r}}.
\label{eq:field}
\ea 
Interestingly, the magnitude of the above field is always smaller than the magnitude of the corresponding field $E_0(r)$ of the 
Maxwell's theory, namely $E(r)<E_0(r)$. In addition $E(r)$ vanishes both at large distances and at the origin (see the 
stationary points  of the potential in Fig. \ref{fig1}). In particular the field has a Coulomb like behavior, $E\sim Q/r$ for $r
\gg l_0$, and is linearly vanishing at short scales, namely $E\approx Qr/l_0^3$ for $r\ll l_0$.  This fact confirms not only  
the good short distance behavior of fields in the presence of T-duality effects, but it also says that, at an intermediate 
distance $0<r_0<\infty$, there exists a maximum for the magnitude of the electric field, namely $E_\mathrm{max}\equiv E(r_0)$.
This can be obtained  as follows:
\ba
0=\frac{dE}{dr}=\frac{Q}{4\pi}\frac{\left(l_0^2-2r^2\right)}{\left(r^2+l_0\right)^{3/2}}\leadsto E_\mathrm{max}=\frac{Q}
{6\sqrt{3}\pi l_0^2}.
\label{eq:maxfield}
\ea
At this point, we recall that the Schwinger limit for a static particle \eqref{eq:SchwingerE} implies that the amount of energy 
stored in its electrostatic field cannot exceed its rest mass. For an electron in standard Maxwell electrodynamics, one has\footnote{We  neglect  multiplicative factors that are irrelevant for the estimate of orders of magnitude.}
\ba
\left(eE_0\right)^2 \lambda_\mathrm{e}^3 <m_\mathrm{e}\leadsto E_0< E_\mathrm{c}\simeq 12 m_\mathrm{e}^2,
\label{eq:schwinlim}
\ea
where the field energy momentum tensor  $T^{00}\sim E^2$ is assumed to be uniformly distributed within the volume $\sim \lambda_
\mathrm{e}^3$, with the electron Compton wavelength and mass being $\lambda_\mathrm{e}$ and $m_\mathrm{e}$ and $\lambda_
\mathrm{e}\sim 1/m_\mathrm{e}$, and $\alpha=e^2\simeq 1/137$ ($1/e\simeq 12$). Eq.  \eqref{eq:schwinlim} actually coincides with 
the Schwinger result \eqref{eq:SchwingerE}.

If one repeats the above calculation with the field modified by the T-duality in \eqref{eq:field}, there will be no substantial 
difference for weak fields. Being $E(r)< E_0(r)$,  if the Schwinger limit is satisfied by Maxwell electrodynamics, 
then T-duality electrodynamics satisfies it too. Conversely, if Maxwell electrodynamics violates the Schwinger limit (namely it 
requires non-linear corrections), this could not be the case for the action \eqref{NLMaxwell05}. It is therefore instructive to 
estimate the value of  $l_0$, at which non-linear  effects set in\footnote{There exist alternative proposals for the value of 
the field strength at which non-linearity become important \cite{RSG73}. Accordingly bounds on $l_0$ can differ.}. To this 
purpose, one can consider the energy density $\sim e^2E^2_\mathrm{max}$ in a volume 
$\sim l_0^3$. From \eqref{eq:schwinlim} one obtains that $l_0 > 2\times 10^{-17}$ m. 

The existence of such a  bound mainly says 
that the scale at which the action \eqref{NLMaxwell05} requires non-linear corrections is around 10 GeV. This sounds surprising because T-duality softens electrostatic  fields. 
If one considers the general requirement $10 \ \mathrm{TeV}< 1/l_0 < \MP$,  it descends that any future observation of the T-duality at scales below (10 TeV)$^{-1}\sim 10^{-22}$m  must be accompanied by non-linearity effects. In other words, there is no energy regime at which T-duality electrodynamics shows up as a linear theory.  
Such a conclusion confirms early results about an interplay between T-duality and 
non-linear field theory actions \cite{Tse96}.

\section{Gauge-invariant calculation}
\label{sec:gaugeinv}

The results up to this point have been obtained by working within a specific gauge, known as Feynman gauge. Electrodynamics is 
gauge invariant. Therefore there is no issue in following a similar method, if the theory is the Maxwell's one. In 
general, one has to make sure that, for a Lagrangian like that in \eqref{NLMaxwell05}, with $\mathcal{O}\neq 1$, the resulting 
potentials \eqref{NLMaxwell45} do not depend on a specific value of the gauge parameter. For sake of consistency, it is 
therefore necessary to repeat the calculation with a gauge-invariant, but path-dependent, variables formalism, proposed by one 
of us in a series of papers \cite{Gaete:1998vr,Gaete:2005nu,Gaete:2007zn}.

To this purpose, we shall first examine the Hamiltonian framework for this theory.  
As a start, it should be noticed that the theory described by (\ref{NLMaxwell05}) contains higher time derivative terms.
In our special case, however, the goal is the study of  static potentials and no particular issue arises. For notational convenience, we will keep the customary notation for the differential operator $\Delta$, even if one can practically substitute it with $ - \nabla ^2$. 

The canonical momenta are ${\Pi ^\mu } = - {\cal O}{F^{0\mu }}$. It is easy to see that $\Pi ^0$ vanishes, we then have the usual constraint equation, which according to Dirac's theory is written as a weak $\left(  \approx  \right)$ equation, that is, $\Pi ^0 \approx 0$. The remaining nonzero momenta must also be written as weak equations, that is, 
%\todo{Piero: Is the canonical momentum $\Pi ^\mu=\frac{\partial {\cal L}}{ \partial \partial_0 A_\mu }$? If I understand correctly, the Hamiltonian density is
%${\cal H}=A_\mu\Pi^\mu -{\cal L}$.  Please explain.}
${\Pi ^i} \approx {\cal O} {F^{i0}}$. The canonical Hamiltonian is now obtained in the usual way via a Legendre transform. It then reads
\begin{equation}
H_\mathrm{C} \approx \int {{d^3}x} \left\{ { - {A_0}{\partial _i}{\Pi ^i} - \frac{1}{2}{\Pi _i}{{\cal O}^{ - 1}}{\Pi ^i} + \frac{1}{4}
{F_{ij}}{\cal O}{F^{ij}}} \right\}, \label{NLMaxwell50}
\end{equation}
 which must also be written as a weak equation. Next, the primary constraint, $\Pi _0  \approx 0$, must be satisfied for all times. An immediate consequence of this is that, using the equation of motion, $\dot Z \approx \left[ {Z,{H_C}} \right]$, we obtain the secondary constraint, 
$\Gamma_1 \left( x \right) \equiv \partial _i \Pi ^i\approx0$. It is easy to check that there are no further constraints in the theory, and that the above constraints are first class.

%\todo{Piero: Do we have $\partial_\mu \Pi^\mu=0$?}

 By proceeding in the same way as in \cite{Gaete:1998vr,Gaete:2005nu,Gaete:2007zn}, we obtain the corresponding total (first class) Hamiltonian that generates the time evolution of the dynamical variables by adding all the first class constraints. We thus write $H = {H_\mathrm{C}} + \int {{d^3}x\left( {{c_0}\left( x \right){\Pi _0}\left( x \right) + {c_1}\left( x \right)\Gamma_{1} \left( x \right)} \right)}$, where ${{c_0}\left( x \right)}$ and ${{c_1}\left( x \right)}$ are arbitrary Lagrange multipliers. 
Now we recall that, when this new Hamiltonian is employed, the equation of motion of a dynamic variable may be written as a strong equation. With the aid of (\ref{NLMaxwell50}), we find that ${\dot A_0}\left( x \right) = \left[ {{A_0}\left( x \right),H} \right] = {c_0}\left( x \right)$, which is an arbitrary function. Since ${\Pi ^0} \approx 0$ always, neither $A^{0}$ nor ${\Pi ^0}$ are of interest in describing this system and may be discarded from the theory. Actually, the term containing $A^{0}$ is redundant, because it can be absorbed by redefining the function $c(x) \equiv c_1 (x) - A_0 (x)$. As a consequence, the Hamiltonian now reads
\begin{equation}
H = \int {{d^3}x} \left\{ {c\left( x \right){\partial _i}{\Pi ^i} - \frac{1}{2}{\Pi _i}{{\cal O}^{ - 1}}{\Pi ^i} + \frac{1}{4}
{F_{ij}}{\cal O}{F^{ij}}} \right\}. %,
 \label{NLMaxwell55}
\end{equation}
%where $c(x) = c_1 (x) - A_0 (x)$.

%\todo{Piero: This is not clear for the reader. What is $c_1(x)$? }

 Now the presence of this new arbitrary function, $c(x)$, is undesirable since we have no way of giving it a meaning in a quantum theory. Hence, according to the usual procedure, we impose a gauge condition such that the full set of constraints becomes second class. A particularly appealing and useful choice is given by
%\todo{Piero: What is a first/second class constraint?}
\begin{equation}
\Gamma _2 \left( x \right) \equiv \int\limits_{C_{\xi x} } {dz^\nu }
A_\nu \left( z \right) \equiv \int\limits_0^1 {d\lambda x^i } A_i
\left( {\lambda x} \right) = 0,     \label{NLMaxwell60}
\end{equation}
where  $\lambda$ $(0\leq \lambda\leq1)$ is the parameter describing the space-like straight path $ z^i = \xi ^i  + \lambda 
\left( {x -\xi } \right)^i $, and $ \xi $ is a fixed point (reference point), on a fixed time slice.
%\todo{Piero: You mean $\xi^i$, dont't you?} 
There is no essential loss of generality if we 
restrict our considerations to $ \xi ^i=0 $. As a consequence, the only nontrivial Dirac bracket is given by
\begin{eqnarray}
\left\{ {A_i \left( x \right),\Pi ^j \left( y \right)} \right\}^ *
&=&\delta{ _i^j} \delta ^{\left( 3 \right)} \left( {x - y} \right) \nonumber\\
&-&\partial _i^x \int\limits_0^1 {d\lambda x^j } \delta ^{\left( 3
\right)} \left( {\lambda x - y} \right). \label{NLMaxwell65}
\end{eqnarray}

To compute the interaction energy we will calculate the 
expectation value of the energy operator $H$ in the physical state $|\Phi\rangle$.
 Let us also mention here that, as was first established by Dirac \cite{Dirac}, the physical states $|\Phi\rangle$ correspond to the gauge invariant ones. Thus, the corresponding physical states take the form
%\todo{Piero: The physical state is defined in terms of $\Psi$ that has not been defined.  Why does the state depend on the charge? I do not follow here. }
\begin{eqnarray}
\left| \Phi  \right\rangle  &\equiv& \left| {\overline \Psi  \left(
\bf y \right)\Psi \left( {\bf y}\prime \right)} \right\rangle \nonumber\\
&=& \overline \psi \left( \bf y \right)\exp \left(
{iQ\int\limits_{{\bf y}\prime}^{\bf y} {dz^i } A_i \left( z \right)}
\right)\psi \left({\bf y}\prime \right)\left| 0 \right\rangle,
\label{NLMaxwell70}
\end{eqnarray}
where $\left| 0 \right\rangle$ is the physical vacuum state and the line integral appearing in the above expression is along a spacelike path starting at ${{\bf y}^ {\prime} }$ and ending at ${{\bf y}}$, on a fixed time slice and $Q$ is the external charge. In this case, each of the states $|\Phi\rangle$ represents a fermion-antifermion pair surrounded by a cloud of gauge fields to maintain gauge invariance.
%\todo{Piero: Changed $q\to Q$.}

 Moreover, using the Hamiltonian formalism developed in \cite{Gaete97}, we have 
\begin{equation}
{\left\{ {{\Pi _k}\left( {\bf x} \right),\Psi \left( {\bf y} \right)} \right\}^ * } = iQ\int_0^1 {d\lambda } {y_k}{\delta ^{\left( 3 \right)}}\left( {{\bf x} - \lambda {\bf y}} \right)\Psi \left( y \right), 
\label{NLMaxwell70-a}
\end{equation}
and 
\begin{equation}
{\left\{ {{\Pi _k}\left( {\bf x} \right),\bar \Psi \left( {\bf y} \right)} \right\}^ * } =  - iQ\int_0^1 {d\lambda } {y_k}{\delta ^{\left( 3 \right)}}\left( {{\bf x} - \lambda {\bf y}} \right)\bar \Psi \left( y \right).
\label{NLMaxwell70-b}
\end{equation}
Next we will consider the state ${\Pi _i}\left( {\bf x} \right)\left| \Phi  \right\rangle$, that is,
\begin{eqnarray}
{\Pi _i}\left( {\bf x} \right)\left| \Phi  \right\rangle \!\!\! &=&\!\!\! \bar \Psi \left( {\bf y} \right)\Psi \left( {{{\bf y}^ \prime }} \right){\Pi _i}\left( {\bf x} \right)\left| 0 \right\rangle   \nonumber\\
&+&\!\!\!\left( {\left[ {{\Pi _i}\left( {\bf x} \right),\bar \Psi \left( {\bf y} \right)} \right]\Psi \left( {{{\bf y}^ \prime }} \right) + \bar \Psi \left( {\bf y} \right)\left[ {{\Pi _i}\left( {\bf x} \right),\Psi \left( {{{\bf y}^ \prime }} \right)} \right]} \right)\left| 0 \right\rangle.
\label{NLMaxwell70-c} 
\end{eqnarray}
By means of (\ref{NLMaxwell70-a}) and  (\ref{NLMaxwell70-b}), we can rewrite  (\ref{NLMaxwell70-c}) in the following way 
%\todo{Piero: Here I guess but I do not understand fully.}
\begin{eqnarray}
\Pi _i \left(\, x\, \right)\left| {\overline \Psi  \left(\, \mathbf{ y}\,
\right)\Psi \left( \,{{\bf y}^ \prime  } \,\right)} \right\rangle  
&=& \overline \Psi  \left( \bf y \right)\Psi \left( {{\bf y}^ \prime }
\right)\Pi _i \left( x \right)\left| 0 \right\rangle \nonumber\\
&+&Q\int_ {\bf y}^{{\bf y}^ \prime  } {dz_i \delta ^{\left( 3
\right)} \left( {\bf z - \bf x} \right)} \left| \Phi \right\rangle. %\nonumber\\
\label{NLMaxwell75}
\end{eqnarray}

 In the case under consideration the expectation value, $\left\langle H \right\rangle _\Phi$, reads
\begin{equation}
\left\langle {{H_\Phi }} \right\rangle  = \left\langle \Phi  \right|\int {{d^3}x}  - \frac{1}{2}{\Pi _i}{{\cal O}^{ - 1}}{\Pi ^i}\left| \Phi  \right\rangle.  \label{NLMaxwell75-a}
\end{equation}

Now making use of equation (\ref{NLMaxwell75}), we find that the expectation value can be brought to the form
\begin{equation}
\left\langle H \right\rangle _\Phi   = \left\langle H \right\rangle
_0 + \left\langle H \right\rangle _\Phi ^{\left( 1 \right)},
\label{NLMaxwell80}
\end{equation}    
where $\left\langle H \right\rangle _0  = \left\langle 0
\right|H\left| 0 \right\rangle$. Whereas the $\left\langle H \right\rangle
_\Phi ^{\left( 1 \right)}$ term is given by
%\todo{Piero: I have to believe you}
\begin{eqnarray}
\left\langle H \right\rangle _\Phi ^{\left( 1 \right)} \!\! &=&\!\!
- \frac{{Q^2 }}{2}
\int {d^3 x} \int_{\bf y}^{{\bf y}^ \prime} {dz_i^ \prime}
\delta ^{\left( 3 \right)}
\left( {{\bf x} - {\bf z}^ \prime} \right) \nonumber\\
\!\!&\times&\!\!\!\left[{- {l_0}\sqrt { - \nabla _x^2} \,{K_1}\left( {{l_0}\sqrt { - \nabla _x^2} } \right)} \right]\int_{\bf y}
^{{\bf y}^ \prime}{dz_i^ \prime} 
\delta ^{\left( 3 \right)}\!\left( {{\bf x} - {\bf z}} \right),
\label{NLMaxwell85}
\end{eqnarray} 
which can also be expressed solely in terms of the new Green's function, $\widetilde G$, namely,\begin{equation}
\left\langle H \right\rangle _\Phi ^{\left( 1 \right)} =  - \frac{{{Q^2}}}{2}\int_{\bf y}^{{{\bf y}^ \prime }} {d{z^{ \prime 
i}}} \int_{\bf y}^{{{\bf y}^ \prime }} {d{z^i}\,\nabla _z^2 \tilde G} \left( {{\bf z},{{\bf z}^ \prime }} \right). 
\label{NLMaxwell90}
\end{equation}

Making use of the foregoing equation
%\todo{Piero: Do you mean the equation for the Green's function? Or just \eqref{NLMaxwell90}?} 
and recalling that the integrals
over ${z^i}$ and ${z^{\prime \,i}}$ are zero except on the contour of
integration, we finally obtain the potential for two opposite charges, located at
${\bf y}$ and ${\bf y^{\prime}}$, 
\begin{equation}
V\left( r \right) =  - \frac{{{Q^2}}}{{4\pi }}\frac{1}{{\sqrt {{r^2} + l_0^2} }},
\label{NLMaxwell95}
\end{equation}
after subtracting a constant term $(\frac{{{Q^2}}}{{4\pi }}\frac{1}{{{l_0}}})$ and $r \equiv |{\bf y} -{\bf y}^{\prime}|$
%\todo{Piero: Changed $L\to r$.}. 

An alternative way of stating the previous result is by considering the expression (see \cite{Gaete:1998vr}):
\begin{equation}
V \equiv Q\left( {\mathcal{A}_0 \left( {\bf y} \right) - \mathcal{A}_0
\left( {\bf y \prime} \right)} \right), \label{NLMaxwell100}
\end{equation}
where the physical scalar potential is given by
\begin{equation}
\mathcal{A}_0 \left( {x^0 ,{\bf x}} \right) = \int_0^1 {d\lambda } x^i
E_i \left( {\lambda {\bf x}} \right), \label{NLMaxwell105}
\end{equation}
with $i=1,2,3$. This equation follows from the vector gauge-invariant field
expression  \cite{Gaete97}
%\todo{As far as I see there are two symbols, namely $\mathcal{A}_\mu $ and $A_\mu$. If $\mathcal{A}_\mu $ is the physical, I do not understand what  $A_\mu$ is. It has to be defined, please. }
\begin{equation}
\mathcal{A}_\mu  \left( x \right) \equiv A_\mu  \left( x \right) +
\partial _\mu  \left( { - \int_\xi ^x {dz^\mu  } A_\mu  \left( z
\right)} \right), \label{NLMaxwell110}
\end{equation}
where the line integral is along a space-like path from the point $\xi$ to $x$, on a fixed slice time. It should be noted that the gauge-invariant variables  (\ref{NLMaxwell110}) commute with the sole first constraint (Gauss law), showing in this way that these fields are physical variables \cite{Dirac}.
We also recall that Gauss' law for the Maxwell case reads $\partial _i \Pi ^i  = J^0$, where we have included the external 
source, $J^{0}$, to represent the presence of ``smeared sources''.
It should be further recalled that the gauge-invariant variables (\ref{NLMaxwell110}) commute with the sole first constraint 
(Gauss' law), corroborating that these fields are physical variables.
 
We also note that for 
\begin{equation}
{J^0}\left( {\bf x} \right) = \frac{{3\sqrt 2 }}{2}\frac{{l_0^2}}{{{\pi ^{{\raise0.5ex\hbox{$\scriptstyle 3$}
\kern-0.1em/\kern-0.15em
\lower0.25ex\hbox{$\scriptstyle 2$}}}}}}\frac{Q}{{{{\left( {{r^2} + l_0^2} \right)}^{{\raise0.5ex\hbox{$\scriptstyle 5$}
\kern-0.1em/\kern-0.15em\lower0.25ex\hbox{$\scriptstyle 2$}}}}}}, \label{NLMaxwell115}
\end{equation} 
the electric field is then
\begin{equation}
E^i  = Q\,\partial ^i \widetilde G \left( x \right), \label{NLMaxwell120}
\end{equation}
where $\widetilde G$ is the Green's function (\ref{NLMaxwell40}).

Finally, replacing this result in (\ref{NLMaxwell105}) and using (\ref{NLMaxwell100}), we readily find that the interaction 
energy for a pair of opposite ``smeared charges'' $Q$, located at ${\bf 0}$ and ${\bf r}$, is given by
 \begin{equation}
V =  - \frac{{{Q^2}}}{{4\pi }}\frac{1}{{\sqrt {{r^2} + l_0^2} }},
\label{NLMaxwell125}
\end{equation}
after subtracting a constant term $(\frac{{{Q^2}}}{{4\pi }}\frac{1}{{{l_0}}})$ and $\left| \mathbf{r} \right| \equiv r$. 

From the above discussion, it has become evident that the understanding of gauge invariance requires a correct identification of physical degrees of freedom of the system. Accordingly, only after such identification has been made, one can legitimately compute the potential by means of the Gauss' law.

\iffalse
It must 
be clear from this discussion that a correct identification of physical degrees of freedom is a key feature for understanding 
the physics hidden in gauge theories.
According to this viewpoint, once that identification is made, the computation of the potential is carried out by means of 
Gauss' law.
\fi

%\section{Conclusions and }
\section{Final remarks}
\label{sec:concl}

In this paper we studied the repercussions of Padmanabhan's propagator in electrodynamics. Such a propagator has the
property of capturing a feature of string theory known as  T-duality. This corresponds to saying that any path integral has to 
be invariant with respect to the exchange of  paths, that result larger and respectively shorter than a fundamental length 
$l_0$. 
Along this line of reasoning, we computed static potentials due to the exchange of virtual photons, by introducing a non-local  
gauge field action for the Padmanabhan's propagator.  

As a main result, we found that the T-duality actually does efficiently work at short scales. Potentials are finite and fields 
are vanishing at short distance. We also discussed possible testbeds of the proposed theory in atomic physics and low energy 
physics experiments like the Casimir capacitor. In contrast to other non-local field theories, like the unparticles, it is 
less realistic to obtain competitive constraints with experiments working at energies below $1/l_0$. This is due to the fact that the proposed theory has an analytic limit to Maxwell electrodynamics for $l_0\to 0$. We also showed that at higher energies  T-duality effects would always show up in combination with non-linear electrodynamics effects.
In agreement with unparicles, T-duality introduces a continuous spacetime dimension with consequent fractalization effects for static potentials. This aspect seems to be related to some aspects of the phenomenology of strange metals and, more in general, to the so called fractional electromagnetism \cite{LLP19}. 

We also presented a double check of the consistency of our approach in relation to the gauge properties of the theory. In particular we proposed an alternative derivation of static potentials based on a gauge-invariant, path-dependent variable formalism.

The topics presented in this work offer one of the rate opportunities for establishing a connection between a quantum gravity modified field theory and a concrete observation of new effects. For this reason, we believe that further investigations have to be done in this direction.

\section*{Acknowledgments}

One of us (P. G.) was partially supported by Fondecyt (Chile) grant 1180178 and by ANID PIA / APOYO AFB180002.
The work of P.N. has partially been supported by GNFM, Italy's National Group for Mathematical Physics.

\end{document}